\documentclass[fleqn,usenatbib]{mnras}

\usepackage{newtxtext,newtxmath}
\usepackage[T1]{fontenc}

\DeclareRobustCommand{\VAN}[3]{#2}
\let\VANthebibliography\thebibliography
\def\thebibliography{\DeclareRobustCommand{\VAN}[3]{##3}\VANthebibliography}

\usepackage{graphicx}	% Including figure files
\usepackage{amsmath}	% Advanced maths commands
\usepackage[dvipsnames]{xcolor}

\newcommand{\eagle}{\textsc{eagle}}

\newcommand{\auriga}{\textsc{auriga}}
\newcommand{\apo}{\textsc{apostle}}

\usepackage{scalerel,tikz}
\usetikzlibrary{svg.path}
\definecolor{orcidlogocol}{HTML}{A6CE39}
\tikzset{orcidlogo/.pic={
 \fill[orcidlogocol] svg{M256,128c0,70.7-57.3,128-128,128C57.3,256,0,198.7,0,128C0,57.3,57.3,0,128,0C198.7,0,256,57.3,256,128z};
 \fill[white] svg{M86.3,186.2H70.9V79.1h15.4v48.4V186.2z}
 svg{M108.9,79.1h41.6c39.6,0,57,28.3,57,53.6c0,27.5-21.5,53.6-56.8,53.6h-41.8V79.1z M124.3,172.4h24.5c34.9,0,42.9-26.5,42.9-39.7c0-21.5-13.7-39.7-43.7-39.7h-23.7V172.4z}
 svg{M88.7,56.8c0,5.5-4.5,10.1-10.1,10.1c-5.6,0-10.1-4.6-10.1-10.1c0-5.6,4.5-10.1,10.1-10.1C84.2,46.7,88.7,51.3,88.7,56.8z};
}}
\newcommand\orcidicon[1]{\href{https://orcid.org/#1}{\mbox{\scalerel*{
\begin{tikzpicture}[yscale=-1,transform shape]
\pic{orcidlogo};
\end{tikzpicture}
}{|}}}}
\title[Effects of subgrid models on angular momentum]{Apostle--Auriga: Effects of stellar feedback subgrid models on the evolution of angular momentum in disc galaxies}

\author[Yang et al.]{Hang Yang $^{1,2,3}$\orcidicon{0000-0003-3279-0134}\thanks{E-mail: hyang@nao.cas.cn},
Shihong Liao $^{1}$\thanks{E-mail: shliao@nao.cas.cn}, 
Azadeh Fattahi $^{4}$,
Carlos S. Frenk $^{4}$,
Liang Gao $^{3,5,1,2}$\thanks{E-mail: lgao@bao.ac.cn},
Qi Guo $^{1,2,5}$,
\newauthor
Shi Shao $^{1,2}$,
Lan Wang $^{1,2}$,
Ruby J. Wright $^{6,7}$,
Guangquan Zeng $^{1,2}$
\\
$^{1}$National Astronomical Observatories, Chinese Academy of Sciences, Beijing 100101, China\\
$^{2}$School of Astronomy and Space Science, University of Chinese Academy of Sciences, Beijing 100049, China\\
$^{3}$School of Physics and Microelectronics, Zhengzhou University, Zhengzhou 450001, China\\
$^{4}$Department of Physics, Institute for Computational Cosmology, University of Durham, South Road, Durham DH1 3LE, UK\\
$^{5}$Institute for Frontiers in Astronomy and Astrophysics, Beijing Normal University, Beijing 102206, China\\
$^{6}$International Centre for Radio Astronomy Research (ICRAR), University of Western Australia, Crawley, WA 6009, Australia\\
$^{7}$Department of Physics, University of Helsinki, Gustaf H{\"a}llstr{\"o}min katu 2, FI-00014 Helsinki, Finland\\
}
\date{Accepted XXX. Received YYY; in original form ZZZ}

\pubyear{2023}

\begin{document}
\label{firstpage}
\pagerange{\pageref{firstpage}--\pageref{lastpage}}
\maketitle

\begin{abstract}
Utilizing the Apostle--Auriga simulations, which start from the same zoom-in initial conditions of Local Group-like systems but run with different galaxy formation subgrid models and hydrodynamic solvers, we study the impact of stellar feedback models on the evolution of angular momentum in disc galaxies. At $z = 0$, {\auriga} disc galaxies tend to exhibit higher specific angular momenta compared to their cross-matched {\apo} counterparts. By tracing the evolution history of the Lagrangian mass tracers of the in-situ star particles in the $z = 0$ galaxies, we find that the specific angular momentum distributions of the gas tracers from the two simulations at the halo accretion time are relatively similar. The present-day angular momentum difference is mainly driven by the physical processes occurring inside dark matter haloes, especially galactic fountains. Due to the different subgrid implementations of stellar feedback processes, {\auriga} galaxies contain a high fraction of gas that has gone through recycled fountain (${\sim} 65$ per cent) which could acquire angular momentum through mixing with the high angular momentum circumgalactic medium (CGM). In {\apo}, however, the fraction of gas that has undergone the recycled fountain process is significantly lower (down to ${\sim} 20$ per cent for Milky Way-sized galaxies) and the angular momentum acquisition from the CGM is marginal. As a result, the present-day {\auriga} galaxies overall have higher specific angular momenta.
\end{abstract}


%%%%%%%%%%%%%%%%%%%%%%%%%%%%%%%%%%%%%%%%%%%%%%%%%%

%%%%%%%%%%%%%%%%% BODY OF PAPER %%%%%%%%%%%%%%%%%%

\section{Introduction}

Angular momentum plays a crucial role in shaping many fundamental properties of galaxies, such as size, kinematics, and morphology \citep[e.g.][]{1980MNRAS.193..189F,1998MNRAS.295..319M,2012ApJS..203...17R, 2015ApJ...804L..40G,2015ApJ...812...29T,2023MNRAS.518.1002R, 2023MNRAS.518.5253Y}{}.
Understanding the evolution of angular momentum is key to comprehending galaxy formation and evolution, especially for the rotation-supported disc galaxies \citep[e.g.][]{1980MNRAS.193..189F, 1998MNRAS.295..319M}{}.
In the standard $\Lambda$ cold dark matter model, galaxies form through cooling and condensation of gas within their host dark matter haloes \citep[]{1978MNRAS.183..341W, 1991ApJ...379...52W}{}.
The initial angular momentum of dark matter haloes is generally acquired through tidal torquing interactions with surrounding inhomogeneities \citep[]{1969ApJ...155..393P, 1970Afz.....6..581D, 1984ApJ...286...38W}{}.
The evolution of angular momentum of dark matter haloes is primarily driven by gravity and thus can be accurately studied using N-body simulations \citep[e.g.][]{2001ApJ...555..240B,2007MNRAS.376..215B,2017ApJ...844...86L}.
However, the angular momentum of galaxies is influenced not only by gravity but also by various complex baryonic processes during their evolution \citep[e.g.][]{2015ApJ...804L..40G,2015ApJ...812...29T,2017ApJ...841...16D,2017MNRAS.470.2262L,2017MNRAS.466.1625Z,2022ApJ...937L..18D,Du2024,2022MNRAS.512.5978R, 2024MNRAS.532.2558Z}, necessitating the use of hydrodynamic simulations to understand the angular momentum evolution of galaxies.

Early hydrodynamical simulations suffered from the so-called `angular momentum catastrophe', where gas experienced over-cooling and drastic angular momentum loss, leading to the formation of compact clumps instead of an extended disc \citep[e.g.][]{1995MNRAS.275...56N,2000ApJ...538..477N,2002MNRAS.335..487M}. Baryonic feedback is considered a necessary heating mechanism to regulate the star formation and prevent over-cooling \citep[see e.g.][for reviews]{2015ARA&A..53...51S, 2017ARA&A..55...59N, 2023ARA&A..61..473C}. However, the limited resolution and the wide range of spatial and temporal dynamic scales in cosmological hydrodynamical simulations make it challenging to include ab initio galaxy formation processes based on first principles. As an alternative method, most baryonic processes have to be modelled using an effective parameterized subgrid approach. The effective stellar feedback is a key ingredient of subgrid physical models for reproducing disc galaxies. In general, the stellar feedback-driven outflows can remove low angular momentum gas, preventing the formation of dense spherical compositions in the early stages of galaxy formation \citep[e.g.][]{2008MNRAS.389.1137S, 2012MNRAS.423.1726S, 2014MNRAS.443.2092U, 2016ApJ...824...79A}. 
With the increased resolution and the adoption of more sophisticated subgrid models, modern hydrodynamical simulations can reproduce more extended disc galaxies \citep[e.g.][]{2010Natur.463..203G,2015ApJ...812...29T}. In addition, baryonic feedback processes are crucial for shaping the observed properties of both circumgalactic and intergalactic media around galaxies \citep[e.g.][]{2015MNRAS.448...59N, 2022MNRAS.514.3113K, Faucher-Giguere2023}. 

As the representatives of the current state-of-the-art cosmological hydrodynamical simulations, both the \textsc{eagle} family \citep[][]{2015MNRAS.450.1937C, 2015MNRAS.446..521S} and the \textsc{illustris} family (including the \textsc{illustris} \citep{2013MNRAS.436.3031V, 2014MNRAS.444.1518V}, \textsc{auriga} \citep{2017MNRAS.467..179G} and \textsc{illustrisTNG} \citep{2018MNRAS.475..624N, 2018MNRAS.473.4077P} simulations)\footnote{The slight difference between the \textsc{auriga} and \textsc{illustrisTNG} models can be found in section 2.3.1 of \cite{2024MNRAS.532.1814G}.} effectively reproduce a large body of observed galaxy properties. For instance, they successfully reproduce diverse morphological types \citep[]{2015MNRAS.454.1886S, 2017MNRAS.472L..45C, 2019MNRAS.487.5416T, 2019MNRAS.483.4140R}{}, the global stellar mass assemble history \citep[]{2015MNRAS.450.4486F, 2015MNRAS.447.3548S, 2017MNRAS.464.1659Q}{}, the colour bimodality distribution \citep[]{2014MNRAS.444.1518V, 2015MNRAS.452.2879T}{}, the galactic size-to-mass relation \citep[]{2017MNRAS.465..722F, 2018MNRAS.474.3976G, 2023MNRAS.518.5253Y}, and so forth.

However, due to the adoption of different subgrid physics models and hydrodynamics solvers, there are discrepancies in galaxy properties between the {\eagle} and \textsc{illustris} families. In particular, \cite{2023MNRAS.518.5253Y} recently revealed significant differences in the correlation between disc galaxy sizes and halo spins in the two simulation families. For Milky Way-sized disc galaxies, both the {\eagle} and \textsc{illustris} families exhibit positive correlations, but the correlation in the former is weaker. Conversely, in the case of dwarf disc galaxies, no correlation is observed in the {\eagle} family, while a positive correlation is identified in the \textsc{illustris} family. This finding raises questions about the validity of classical semi-analytical disc formation models \citep[e.g.][]{1998MNRAS.295..319M,2011MNRAS.413..101G}, which predict the positive correlation between disc size and halo spin. It becomes crucial to investigate the reasons behind the distinct stellar disc evolution (especially the evolution of angular momentum) resulting from the two types of subgrid models in these simulations.

In this work, we further investigate the formation histories of disc galaxies simulated by the {\eagle} and \textsc{illustris} families, with a primary focus on examining the impact of different stellar feedback implementations on the angular momentum evolution of these simulated disc galaxies. We utilize two suites of high-resolution zoom-in simulations, {\apo} \citep[]{2016MNRAS.457.1931S,2016MNRAS.457..844F} and {\auriga} \citep[]{2017MNRAS.467..179G}, derived from the {\eagle} and \textsc{illustris} families, respectively. Specifically, for the {\auriga} suite, we adopt the simulations started from the same initial conditions as the {\apo} AP-V1 and AP-S5 runs but run with the {\auriga} model \citep{2022MNRAS.514.3113K}. This approach enables a direct matching of haloes/galaxies between the two families, facilitating an apples-to-apples comparison. 

By selecting corresponding disc galaxies within the matched halo pairs and employing the mass element tracer technique, we perform a detailed particle-based Lagrangian analysis to compare the angular momentum evolution between the two simulation families.

This paper is organized as follows. In Section \ref{sec:sim}, we provide the details of the two simulations and the galaxy sample selection. 
The method to trace the evolution of particles' angular momenta is described in Section \ref{sec:method}. 
We present our results in Section \ref{sec:results} and summarize in Section \ref{sec:summary}.

\section{Simulations and galaxy samples}\label{sec:sim}

\subsection{Simulation overview}

This work makes use of two suites of zoom-in hydrodynamical simulations representing Local Group-like volumes. The first set comprises two volumes, namely AP-V1 and AP-S5, selected from the {\apo} project \citep[]{2016MNRAS.457.1931S,2016MNRAS.457..844F}. The {\apo} Local Group-like regions are drawn from the dark matter-only simulation DOVE according to their mass and kinematics at present. For detailed information on the selection conditions, we direct the interested reader to \cite{2016MNRAS.457..844F}. The zoom-in initial conditions (ICs) are created by \textsc{ic\_gen} \citep[]{2010MNRAS.403.1859J}{} which employs the second-order Lagrangian perturbation theory. 

The {\apo} simulations are run using a highly modified version of the \textsc{gadget-3} code, which is an improved version of \textsc{gadget-2} \citep[]{2005MNRAS.364.1105S}{}. The gas properties are calculated with the pressure-entropy formulation of smoothed particle hydrodynamics (SPH) method \citep[]{2013MNRAS.428.2840H,2015MNRAS.454.2277S}{}. All the numerical set-up and parameters are the same as the {\textsc{eagle}} reference model \citep{2015MNRAS.450.1937C, 2015MNRAS.446..521S}.

The second suite of simulations use the same ICs as the {\apo} suite, but is run with the {\auriga} galaxy formation model \citep[]{2017MNRAS.467..179G}. The {\auriga} simulations are performed with the moving-mesh magneto-hydrodynamics code \textsc{arepo} \citep[]{2010MNRAS.401..791S}{}. For convenience, we still call this suite of simulations as {\auriga}, but note that it adopts different ICs compared to the original {\auriga} project \citep[]{2017MNRAS.467..179G,Grand2024}{}.

Both two simulation sets adopt the WMAP-7 cosmological parameters \citep[]{2011ApJS..192...18k} which are $\Omega_{\Lambda}=0.728$, $\Omega_{\rm m}=0.272$, $\Omega_{\rm b}=0.0455$, $h=0.704$,  $\sigma_{8}=0.81$, and $n_{\rm s}=0.967$. The initial gas (dark matter) particle/cell mass is about $1.2(5.9)\times10^5 {\rm M}_{\sun}$, and the maximum softening length is 307 pc in both simulations. The Friends-of-Friends \citep[FOF,][]{1985ApJ...292..371D} and SUBFIND \citep[]{2001MNRAS.328..726S, 2009MNRAS.399..497D} algorithms
are applied to identify halo and subhalo structures, respectively. In this work, the halo virial radius, $R_{200}$, is defined as the radius within which the mean density is 200 times the critical density of the Universe. Unless otherwise specified, we use the star particles within the 3D aperture with a radius of $0.1 R_{200}$ to compute the stellar properties. Both simulations contain $128$ particle snapshots and corresponding group catalogs, with a time interval being about 100 Myr. We match the dark matter particles within subhaloes by particle IDs to identify the main progenitor branch of the merger tree across different snapshots. Specifically, for a subhalo in snapshot $S_{n}$, we consider its dark matter particles within twice the half-total-mass radius, and define the subhalo in the previous snapshot (i.e. snapshot $S_{n-1}$) that contains the most matched particles as the main progenitor.

Both {\apo} and {\auriga} simulations use the TreePM method to compute gravity. However, they adopt different hydrodynamical solvers (i.e. SPH for the former versus moving-mesh for the latter), which can potentially introduce differences in the baryon cycle \citep[e.g.][]{Keres2012,2013MNRAS.429.3353N}. Nevertheless, we expect that the effects of hydrodynamical solvers are secondary compared to those of the baryonic subgrid models, which are elaborated in the next subsection, especially for the evolution of gas within the halo virial radius \citep[see][]{2012MNRAS.423.1726S, 2014MNRAS.442.1992H, 2015MNRAS.454.2277S,2018MNRAS.480..800H}. As we will show in Section~\ref{subsec:overall_evo_am}, the distributions of gas angular momenta exhibit noticeable similarities in both simulations when the gas enters the halo virial radius, with larger differences emerging gradually thereafter. 

\subsection{Subgrid models}

Both {\apo} and {\auriga} simulations incorporate key baryonic processes relevant to galaxy formation, such as radiative cooling \citep[]{2009MNRAS.393...99W}{}, star formation \citep[]{2008MNRAS.383.1210S}{}, stellar evolution and metal enrichment \citep[]{2009MNRAS.399..574W}{}, stellar feedback \citep[]{2012MNRAS.426..140D, 2013MNRAS.436.3031V}{}, and black hole mass growth and active galactic nuclei (AGNs) feedback \citep[][]{2005MNRAS.361..776S,2009MNRAS.398...53B,2015MNRAS.454.1038R}. The detailed descriptions of the subgrid models used in {\apo} and {\auriga} can be found in \citet{2015MNRAS.446..521S} and \citet{2017MNRAS.467..179G}, respectively. In this subsection, we mainly summarize the differences in the implementation of stellar feedback, which plays a significant role in influencing the evolution of baryonic angular momentum in the dwarf and Milky Way-sized galaxies studied in this work.

In {\auriga}, a gas cell is assumed to be star-forming and thermally unstable when its density is higher than the threshold density of $n_{\rm th} = 0.13~\rm{cm^ {-3}}$. Once entering the thermally unstable star-forming regime, this gas cell is stochastically converted into a star particle according to a probability that scales exponentially with time in units of the star-forming time-scale $t_{\rm sf} = 2.2~{\rm Gyr}$. In {\apo}, the star formation is also implemented stochastically, but following the pressure-dependent Kennicutt-Schmidt law \citep{2008MNRAS.383.1210S} with a metallicity-dependent density threshold value $n_{\rm th} = 0.1 (Z/0.002)^{-0.64} \rm{cm^{-3}}$, where $Z$ is the gas metallicity (i.e. the mass fraction of gas in elements heavier than helium). Although the density threshold has some differences, the star formation processes are mainly self-regulated by feedback \citep[e.g.][]{2010MNRAS.402.1536S}.

In both {\apo} and {\auriga}, a star particle represents a single stellar population (SSP), and the \citet[]{2003PASP..115..763C} initial mass function is adopted to specify the distribution of stellar masses contained in an SSP. Hence, the amount of supernovae (SNe) energy per unit stellar mass released into surrounding gas is similar in two simluations. However, the SNe feedback implementation in the two simulations are different: 

In {\auriga}, the kinetic SNe feedback is implemented by converting a gas cell into a wind particle \citep[]{2013MNRAS.436.3031V}. The wind particle is kicked with an isotropically random direction, and the wind speed $v_{\rm{wind}}$ is proportional to the local 1D dark matter velocity dispersion $\sigma_{\rm{dm}}$ which is calculated from 64 nearest dark matter particles, $v_{\rm{wind}} = 3.46\sigma_{\rm{dm}}$. After launch, a wind particle is decoupled from the hydrodynamical computation until either reaching a low-denisty region (5 per cent of star-forming threshold density) or exceeding a maximum time (2.5 per cent of the Hubble time at launch). Once the wind particle recouples, it deposits its mass, metals, momentum, and energy into the intersected gas cells. 

In contrast, the {\apo} simulations adopt the thermal implementation of stellar feedback \citep[]{2012MNRAS.426..140D}. Specifically, the SNe energy of an SSP is stochastically ejected to a few neighboring particles within the SPH kernel size, resulting in an increase in the internal energy of these gas particles. Note that in {\apo}, the stellar feedback is performed only through the thermal mechanism. To avoid that the cooling time is shorter than the sound cross time, each heated gas particle is subject to the same high enough temperature increase, $\Delta T_{\rm SN} = 10^{7.5}$ K. As demonstrated by \citet{2012MNRAS.426..140D}, such a sufficiently high temperature boost is required for the injected thermal energy to be efficiently converted into kinetic energy, leading to massive, large-scale outflows.

\subsection{Galaxy samples}\label{subsec:sample}

\begin{figure*}
	\includegraphics[width=\textwidth]{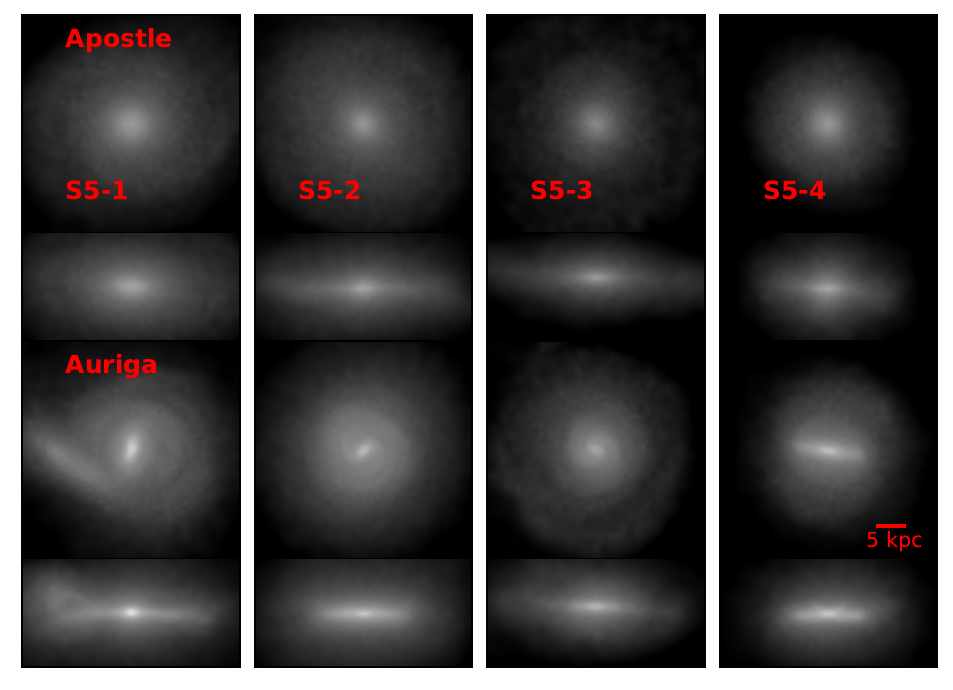}
	\caption{Visualization of the stellar components of galaxies at $z=0$. The top two rows show face-on and edge-on projections of {\apo} galaxies, while the bottom two rows are for {\auriga} galaxies. From left to right, the S5-1 to S5-4 pairs are plotted.}
    \label{fig:project_1}
\end{figure*}

\begin{figure*}
	\includegraphics[width=\textwidth]{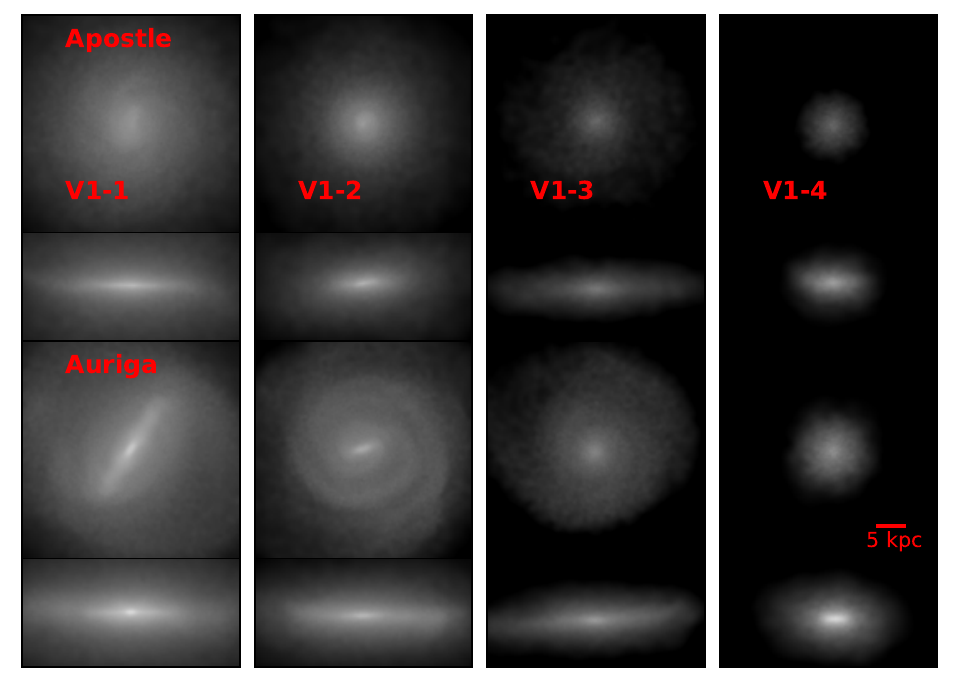}
	\caption{Similar as Figure~\ref{fig:project_1}, but for the V1-1 to V1-4 pairs.}
    \label{fig:project_2}
\end{figure*}

In this work, we focus our analysis on central disc galaxies, which are classified by applying a criterion based on the kinematics-based morphology parameter $\kappa > 0.5$ \citep{2023MNRAS.518.5253Y}. Here, following \citet{Sales2012}, the $\kappa$ parameter is defined as the ratio between the rotational energy, $K_{\rm rot}$, and the total kinetic energy, $K$, i.e.

\begin{equation}
 \kappa= \frac{K_{\rm rot}}{K}=\frac{\sum_i{(1/2)}m_i [(\hat{\boldsymbol{L}}\times{\hat{\boldsymbol{r}}_i})
 \cdot{\boldsymbol{v}_i}]^2}
{\sum_i{(1/2)}m_i{v_i^2}},
\label{equ:Kappa}
\end{equation}
where $\hat{\boldsymbol{L}}$ is the unit vector of the total stellar angular momentum, $\hat{\boldsymbol{r}}_i$ is the unit vector of the position vector from the galaxy centre for star particle $i$, $\boldsymbol{v}_i$ is its velocity vector in the centre of mass frame, and $m_i$ is its particle mass. For each galaxy at $z=0$, $\kappa$ is computed using all star particles within $0.1 R_{200}$. 

In total, we identify 12 isolated disc galaxies from the high resolution regions of two {\auriga} runs. To ensure reliable resolution, we select only the four most massive isolated disc galaxies from each {\auriga} run (8 galaxies in total), each containing at least 50000 star particles. The subhalo IDs, virial radii ($R_{200}$), and virial masses ($M_{200}$) of the eight selected {\auriga} galaxies are provided in Columns 2--4 of Table~\ref{tab:sample}. The galaxies from the two volumes are denoted as S5-1 to S5-4 and V1-1 to V1-4, respectively.

To offer an apples-to-apples comparison, we further match the eight $z=0$ counterparts from the {\apo} runs by comparing the virial radii and galaxy centres (i.e. the position of the particle with the minimum gravitational potential energy, $\mathbfit{r}_{\rm min}$). Specifically, for each {\auriga} galaxy, we loop over all {\apo} galaxies and compute the relative differences,
\begin{equation}
    \Delta_{R} = \frac{\left|R^{\rm{{\apo}}}_{200}-R^{\rm{{\auriga}}}_{200}\right|}{R^{\rm{{\auriga}}}_{200}}
\end{equation}
and
\begin{equation}
    \Delta_{r} = \frac{\left|\mathbfit{r}_{\rm min}^{\rm {\apo}} - \mathbfit{r}_{\rm min}^{\rm {\auriga}}\right|}{R_{200}^{\rm {\auriga}}}.
\end{equation}
The matched {\apo} galaxy is defined as the one with the minimum value of $\Delta=\sqrt{\Delta_R^2+\Delta_r^2}$. We have further verified the cross-matching results by comparing the dark matter particle IDs from the matched galaxy pairs. The FOF group IDs, the SUBFIND group IDs, the virial radii, and the virial masses of the eight {\apo} galaxies can be found in Columns 5--8 of Table~\ref{tab:sample}.
%To do this, we compute the relative differences of virial radii and galaxy centres. We have also tested with an alternative approach by cross-matching dark matter particle IDs, and the results are the same.

To distinguish the {\auriga} and {\apo} galaxies, we introduce the prefixes `Au-' and `Ap-' to the galaxy names, respectively. For example, `Au-S5-1' and `Ap-S5-1' represent the most massive galaxy pairs within the S5 volume in the two respective simulations. The baryonic properties of the galaxy pairs are summarized in Table~\ref{tab:disk}.

The {\auriga} galaxies tend to have slightly higher $M_{200}$ ($R_{200}$) compared to the {\apo} counterparts, with the relative differences being $\la 20 \%$ ($\la 10 \%$), as a result of the higher baryonic fractions in {\auriga} (see Table~\ref{tab:disk}).
% \zeng{\textit{[I prefer to describe it as: "As shown in Table 2, the Auriga galaxies have slightly higher M200, R200, and baryonic fractions compared to their Apostle counterparts."
% % The original description used "... as a result of ...," which, to me, conveys a somewhat weird logic.
% ]}}

% Our selected galaxies span from dwarf to Milky Way-sized systems,

In this study, we analyze a total of 8 pairs of disc galaxies spanning from dwarf to Milky Way-sized systems, 
i.e. with their present-day virial masses $M_{200}$ ranging from $6.1 \times 10^{10}$ to $1.6 \times 10^{12} {\rm M}_{\sun}$ and virial radii raning from ${\sim} 80$ to ${\sim} 250$ kpc. The primary motivation of this study is to examine the influence of different stellar feedback implementations on the evolution of angular momentum. For Milky Way-sized galaxies, the comparison can be partially contaminated by the presence of AGN feedback. By incorporating a sample of dwarf galaxies, which do not harbour supermassive black holes, we can disentangle the impact of AGN feedback from our overall conclusions. 
% {\color{green}GQ: you donot discribe the selection of the pairs of dwarf galaxies and in Talbe 1 only MW-sized galaxies are included. I am a bit confused which galaxy pairs are used. } 

To visually inspect the stellar components of the selected galaxies, we plot the face-on and edge-on projected maps of the matched galaxy pair from the volumes S5 and V1 in Figures \ref{fig:project_1} and \ref{fig:project_2}, respectively. The face-on direction is determined by the direction of the stellar specific angular momentum (sAM) computed using star particles within $2 r_{*, 1/2}$. \footnote{To mitigate the influence of the merging satellite in Au-S5-1 (see Figure~\ref{fig:project_1}) and galactic disc warps in other runs, we determine the face-on direction by adopting the stellar sAM within a radius of $2 r_{*, 1/2}$ instead of the aperture radius of $0.1R_{200}$.} As illustrated in Figures \ref{fig:project_1} and \ref{fig:project_2}, all samples exhibit distinct disc structures. In {\auriga}, galaxies display a higher central stellar surface density and a more complex dynamic structure, including bars and spiral arms. This phenomenon may originate from the non-axisymmetric instability as a result of the higher baryonic mass to dark matter mass ratio \citep[]{1973ApJ...186..467O, 1982MNRAS.199.1069E}{} in {\auriga} compared to {\apo}.

\begin{table*}
	\centering
	\caption{The properties of the matched galaxy pairs at $z=0$. From left to right, the columns are: (1) Galaxy name;  (2) Subhalo ID in {\auriga}; (3) Galaxy virial radius in {\auriga}; (4) Galaxy virial mass in {\auriga}; (5) Group ID in {\apo}; (6) Subgroup ID in {\apo}; (7) Galaxy virial radius in {\apo}; (8) Galaxy virial mass in {\apo}. }
    % {\color{green}GQ: I'd suggest to keep only the pair of galaxies used for this analysis. It is better to use GroupID instead of Group Number for the 5th colomn, as well as for 'subgroupnumber'.}}
	
	\label{tab:sample}
	\begin{tabular}{lccccccc} % c columns alignment for each
		\hline
		Name & SubhaloID & $R_{200}$ [kpc] & $\log_{10} M_{200}$ [${\rm M}_{\sun}$] & GroupID & SubGroupID &
		  $R_{200}$ [kpc] & $\log_{10} M_{200}$ [${\rm M}_{\sun}$]  \\
		& ({\auriga}) & ({\auriga}) & ({\auriga}) & ({\apo}) & ({\apo}) & ({\apo}) & ({\apo}) \\
		\hline
		S5-1 & 0 & 205 & 11.99 & 1 & 0 & 199 & 11.96 \\
		S5-2 & 344 & 191 & 11.90 & 1 & 1 & 186 & 11.89 \\
		S5-3 & 562 & 149 &  11.58 & 2 & 0 & 141 & 11.51 \\
		S5-4 & 730 & 165 & 11.72 & 3 & 0 & 158 & 11.66 \\
		
		V1-1 & 0 & 242 & 12.22 & 1 & 0 & 242 & 12.21 \\
		V1-2 & 355 & 213 & 12.05 & 2 & 0 & 207 & 12.01 \\
		V1-3 & 843 & 110 & 11.19 & 4 & 0 & 103 & 11.10 \\
		V1-4 & 869 & 87 & 10.89 & 7 & 0 & 80 & 10.78 \\
		\hline  
	\end{tabular}
\end{table*}

\begin{table*}
	\centering
	\caption{The baryonic properties of the matched galaxies at $z=0$. From left to right, the columns are: (1) Galaxy name; (2) Stellar mass within $0.1R_{200}$ (within $R_{200}$); (3) Gas mass within $0.1R_{200}$ (within $R_{200}$); (4) Baryon fraction within $0.1R_{200}$ (within $R_{200}$); (5) Half stellar mass radius; (6) Stellar specific angular momentum within $0.1R_{200}$; (7) Kinematics-based morphology parameter; (8) Ex-situ stellar mass fraction. }
	
	\label{tab:disk}
	\begin{tabular}{lcccccccc} % c columns alignment for each
		\hline
		Name & $M_{*}$  & $M_{\rm gas}$ & $f_{\rm b}$ & $r_{*, 1/2}$  & $j_*$  & $\kappa$ & $f_{\rm ex}$  \\
             & [$10^{10} {\rm M}_{\sun}$] & [$10^{10} {\rm M}_{\sun}$] & & [kpc] & [kpc km s$^{-1}$] & & & \\
		\hline
		Au-S5-1 & 5.49 (5.70) & 2.63 (8.21) & 0.082 (0.141) & 3.0 & 676 & 0.63 & 0.12 \\
    	Ap-S5-1 & 1.95 (2.11) & 0.70 (4.86) & 0.029 (0.076) & 4.0 & 298 & 0.61 & 0.05 \\
		\hline
		Au-S5-2 & 3.82 (4.03) & 1.64 (6.13) & 0.068 (0.127) & 4.6 &  699 & 0.65 & 0.04 &  \\
  		Ap-S5-2 & 1.40 (1.52) & 0.57 (5.41) & 0.025 (0.090) & 4.9 & 479 & 0.57 & 0.10  \\
		\hline
		Au-S5-3 & 1.74 (1.93) & 1.21 (4.45) & 0.078 (0.168) & 3.7 & 455 & 0.67 & 0.05 \\
		Ap-S5-3 & 0.63 (0.77) & 0.17 (1.81) & 0.025 (0.080) & 4.5 & 350 & 0.58 & 0.08 \\
	   	\hline
        Au-S5-4 & 3.22 (3.31) & 1.09 (3.76) & 0.082 (0.135) & 2.7 & 410 & 0.53 & 0.02 \\
		Ap-S5-4 & 1.11 (1.16) & 0.44 (2.31) & 0.034 (0.076) & 3.3 & 214 & 0.51 & 0.05 \\
  		\hline
		Au-V1-1 & 10.25 (11.01) & 3.09 (11.55) & 0.081 (0.137)  & 5.7 & 1295 & 0.59 & 0.08  \\
		Ap-V1-1 & 4.73 (5.39) & 1.94 (16.21) & 0.041 (0.132) & 6.7 &  1110 & 0.68 & 0.15 \\
		\hline
		Au-V1-2 & 3.56 (3.83) & 2.18 (8.81) & 0.051 (0.113) & 7.5 & 1129 & 0.70 & 0.12  \\
		Ap-V1-2 & 1.85 (2.00) & 0.75 (6.98) & 0.025 (0.088) & 3.9 & 76 & 0.53 & 0.12 \\
		\hline
  		Au-V1-3 & 0.62 (0.80) & 0.49 (2.23) & 0.071 (0.195) & 6.2 & 464  & 0.70 & 0.16 \\
		Ap-V1-3 & 0.17 (0.27) & 0.06 (0.84) & 0.019 (0.088) & 6.9 & 426  & 0.62 & 0.16 \\
		\hline
		Au-V1-4 & 0.43 (0.44) & 0.29 (0.76) & 0.092 (0.154) & 1.6 & 71  & 0.50 & 0.01 \\
		Ap-V1-4 & 0.08 (0.08) & 0.03 (0.10) & 0.018 (0.030) & 2.7 & 76  & 0.49 & 0.01 \\
		\hline  
	\end{tabular}
\end{table*}

%\section{Methods}\label{sec:method}
\section{Angular momentum tracing method}\label{sec:method}

The Lagrangian point of view has been extensively utilized in previous simulation studies to examine the evolution of galaxy angular momentum \citep[e.g.][]{1984ApJ...286...38W,Sugerman2000,Porciani2002,2017ApJ...841...16D,2017MNRAS.470.2262L,2023ApJ...943..128S}. This approach involves tracing the mass elements comprising a $z=0$ structure back to earlier snapshots and computing their angular momentum to construct the evolutionary history. It has proven to be insightful in understanding the origin and evolution of galaxy angular momentum, especially in modern galaxy formation simulations \citep[e.g.][]{2017ApJ...841...16D}. In this study, we also employ this approach, and the detailed methodology is described as follows.

\subsection{Particle tracing}

To understand how the angular momentum of the stellar disc evolves into the final state, we trace back the history of each star particles within $0.1 R_{200}$ in the galaxy at $z=0$.
In {\apo}, the Lagrangian SPH solver is used, and each star particle is converted from a corresponding progenitor gas particle. Therefore, it is natural to directly track a mass element from its very beginning as a gas particle to its final state as a star particle by using its unique particle ID.

However, in {\auriga}, the gas `particles' correspond to mesh cells rather than actual Lagrangian particles in SPH. Although {\sc arepo} uses a quasi-Lagrangian moving mesh scheme that minimizes mass fluxes between cells, there is still non-zero mass exchange between cells. Furthermore, the number of cells is not conserved in order to adjust the mass resolution when the refinement (de-refinement) operation is used to split (merge) gas cells. Therefore, each gas cell might not represent the same material elements during the simulation. The additional tracer particle must be adopted to track the mass flow. The {\auriga} simulations include Monte Carlo tracer particles \citep[]{2013MNRAS.435.1426G} to track the mass flow throughout the simulation in a Lagrangian way. Each gas cell contains one tracer particle at the beginning of the simulation. According to mass exchange or mesh refinement (de-refinement), tracer particles are then exchanged between neighbouring cells in a probabilistic way. Tracer particles can also be retained when their host gas cells have been transferred to star particles. In this paper, we select all tracer particles resided in the selected stellar composition of the galaxy at $z=0$ in {\auriga}, and then record the information of their corresponding host particles at each snapshot. For convenience, we collectively refer to both SPH gas particles in {\apo} and Monte Carlo tracers in {\auriga} as `gas tracers' in the following paragraphs.

\subsection{Characterizing the evolution of angular momentum}\label{subsec:chara_events}

Once we have the evolutionary history of each $z=0$ star particle within $0.1 R_{200}$, we know whether it has an {\it in-situ} (formed within the main progenitor branch of the merger tree) or {\it ex-situ} (formed outside) origin. Since our focus is on how the baryonic feedback from the target galaxy affects its angular momentum evolution, we analyze only the in-situ star particles and disregard those originating from ex-situ accretion. Table~\ref{tab:disk} presents the ex-situ stellar mass fractions, $f_{\rm{ex}}$, in the rightmost column. Notably, for our galaxy samples, the majority of star particles have an in-situ origin (i.e. $f_{\rm ex} \la 15\%$). Note that similar considerations are also adopted in \citet{2017ApJ...841...16D} when studying the impact of galactic winds on stellar angular momentum in the {\textsc{illustris}} simulations and their ex-situ mass fractions are similar, i.e. ${\sim} 10\%$.
% \zeng{\textit{[I think if it has been previously mentioned that (almost) all galaxy properties are measured within 0.1R200, then it is unnecessary to repeatedly specify "within 0.1R200" later on.]}}

For each tracing particle $i$, we compute its sAM at each snapshot,

\begin{equation}
    \mathbfit{j}_i = (\mathbfit{r}_i - \mathbfit{r}_{\rm min}) \times (\mathbfit{v}_i - \mathbfit{v}_{\rm c}),
\end{equation}
where $\mathbfit{r}_i$ and $\mathbfit{v}_i$ are the position and velocity of particle $i$, $\mathbfit{r}_{\rm min}$ is the galaxy centre, and $\mathbfit{v}_{\rm c}$ is the centre of mass velocity computed from all particles in the galaxy (including dark matter, gas, stars, and black holes if available). The $z$-axis component of sAM, $j_{i,z}$, is computed by projecting $\mathbfit{j}_{i}$ onto the face-on direction defined at $z = 0$ (see Section~\ref{subsec:sample} for details).

The gas tracers we analyzed in this paper eventually fall into the gas disc and form stars. However, after entering the disc and crossing the star formation threshold for the first time, not all gas remains in this state and subsequently converts into star particles. Some gas is ejected from the high-density ISM region, temporarily leaves the star-forming state, and later returns. This phenomenon is known as galactic fountains \citep[see][for a review]{2017ASSL..430..323F}. This motivates us to compare the sAM of particles from the {\apo} and {\auriga} simulations at the following characteristic time \citep{2017ApJ...841...16D}, as we quantify the evolution of galaxy angular momentum:

\begin{enumerate}
\item {\it Halo accretion time} ($z_{\rm{acc}}$): the tracer belongs to the main progenitor of the halo for the first time.

\item {\it First star-forming time} ($z_{\rm{fsf}}$): the gas element enters the central galaxy disc (the distance to the galaxy centre is less than $0.1R_{200}$) and crosses the star-forming density threshold for the first time. 

\item {\it Last star-forming time} ($z_{\rm{lsf}}$): the gas element crosses the star-forming density threshold for the last time. Note that for gas tracers that have never exited the star-forming state before being converted into stars, their $z_{\rm{fsf}}$ and $z_{\rm{lsf}}$ are identical.

\item {\it Star formation time} ($z_{\rm{sf}}$): the tracer is converted from the gas phase into the stellar phase.

\item {\it The present time} ($z = 0$): the final snapshot of simulations.
\end{enumerate}

%\textcolor{red}{The gas we analyzed in this paper eventually falls into the gas disc and forms stars. However, after entering the disc and crossing the star formation threshold for the first time, not all gas remains in this state and subsequently converts into star particles. For gas tracers that have never leave the star-forming state before be converted into stars, they have the same $z_{\rm{fsf}}$ and $z_{\rm{lsf}}$. Some gas is ejected from the high-density ISM region, temporarily leaves the star-forming state, and later returns. This phenomenon is known as galactic fountains \citep[see][for a review]{2017ASSL..430..323F}.}   

\section{Results}
\label{sec:results}
\subsection{Angular momentum properties at $z=0$}
\begin{figure}
	\includegraphics[width=\linewidth]{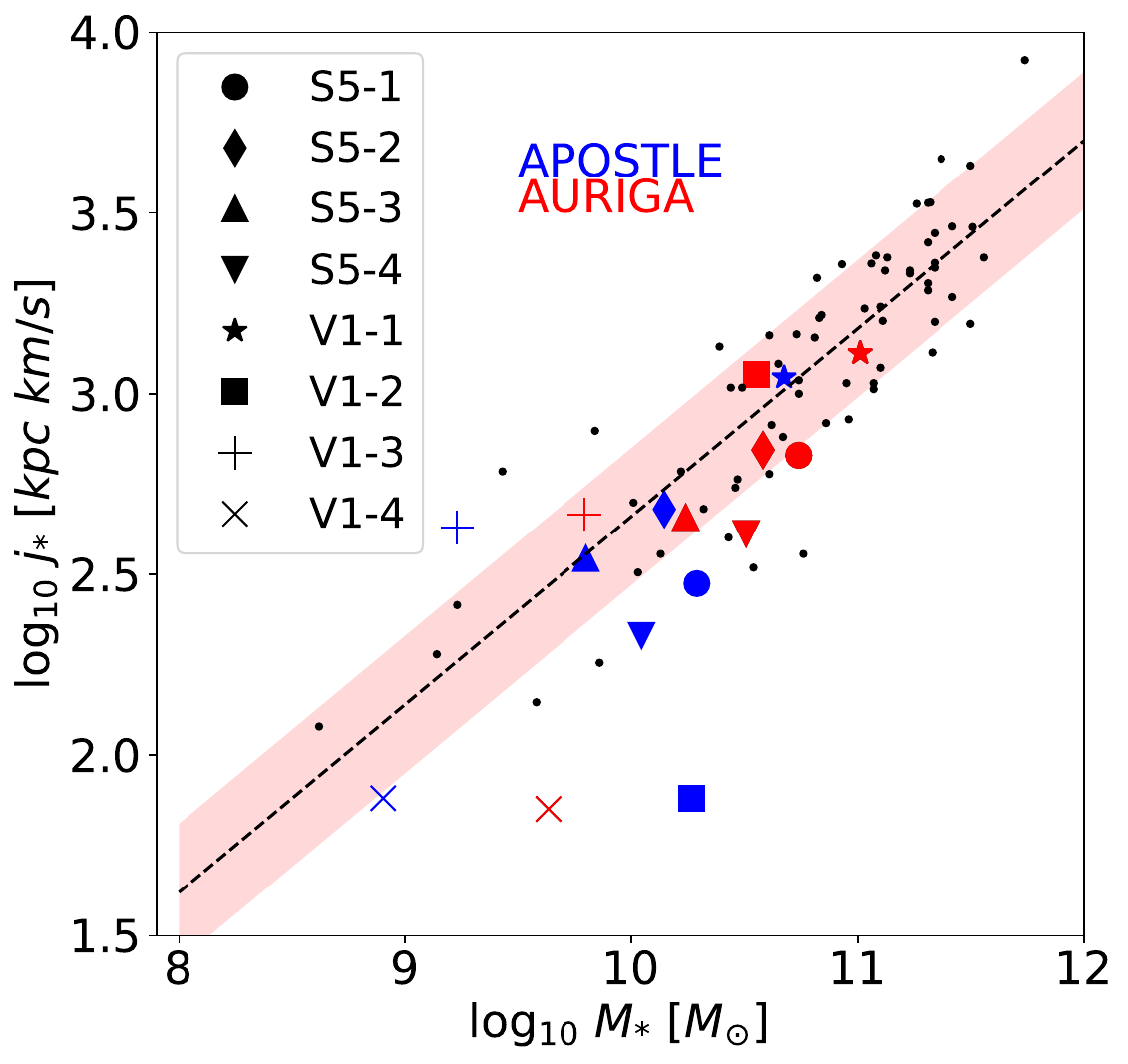}
        \caption{Stellar $j_{*}$--$M_{*}$ relation at $z=0$. Here $j_{*}$ and $M_{*}$ are computed using the star particles within $0.1 R_{200}$. Different markers are used to distinguish different matched galaxy pairs from the {\auriga} (red) and {\apo} (blue) simulations. The little black dots show the observed spiral galaxies from \citet{2012ApJS..203...17R} (the `All spirals, total, intrinsic' sample), and the dashed line and shaded region represent the corresponding best-fitting relation and $1\sigma$ scatter. Both {\auriga} and {\apo} galaxies follow the observed $j_{*}$--$M_{*}$ relation, but {\auriga} galaxies tend to exhibit slightly higher sAM and stellar masses compared to their {\apo} counterparts. 
        }
    \label{fig:jM}
\end{figure}

To compare the stellar angular momentum properties at $z = 0$ and check if both simulations agree with observations, in Figure~\ref{fig:jM}, we plot the $z = 0$ stellar sAM--mass ($j_{*}$--$M_{*}$) relation of the simulated galaxies together with the observations from \citet{2012ApJS..203...17R}. Overall, both {\auriga} and {\apo} simulations reasonably reproduce the observed relation. This is consistent with the recent results of \citet{Hardwick2023}, who found that the stellar $j_{*}$--$M_{*}$ relations from both {\textsc{eagle}} and IllustrisTNG simulations agree well with the xGASS observations \citep{Hardwick2022}. It also echos previous literature that studies the $j_{*}$--$M_{*}$ relation in either {\textsc{eagle}} \citep[e.g.][]{Lagos2017} or IllustrisTNG \citep[e.g.][]{2022ApJ...937L..18D,2022MNRAS.512.5978R,Fall2023} simulations.

We also notice that {\auriga} galaxies tend to exhibit marginally higher stellar mass and sAM compared to their {\apo} counterparts. This trend aligns with the slightly higher rotation-to-kinetic energy ratios observed in {\auriga} galaxies (see the measurement results of $\kappa$ parameters in Table~\ref{tab:disk}). What causes the {\auriga} galaxies to show relatively higher stellar angular momentum at $z=0$? Does this difference already appear in the initial gas accretion, or does it gradually develop due to later baryonic processes occurring inside haloes? To answer these questions, in the next subsection, we trace the Lagrangian tracers back to $z = z_{\rm acc}$ and compare the sAM distributions in the two simulations.

\subsection{Distribution of trace-back angular momentum at $z_{\rm acc}$}
\label{subsec:overall_evo_am}

\begin{figure*}
	\includegraphics[width=\textwidth]{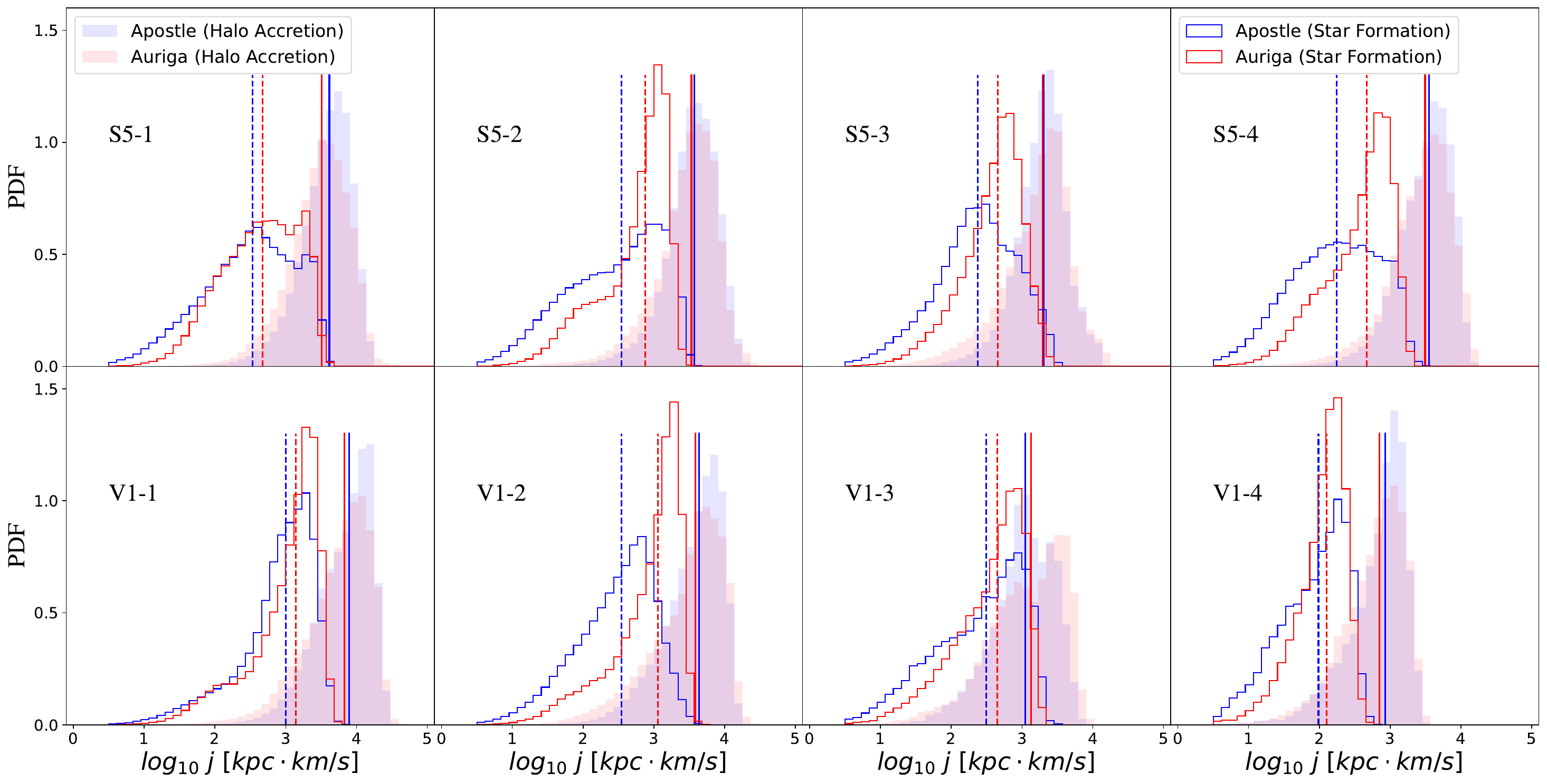}
        \caption{Probability distribution functions of the sAM magnitude ($j$) for the tracers when they first enter the main halo progenitor (at $z = z_{\rm acc}$, filled) and when they are converted into stars (at $z = z_{\rm sf}$, unfilled). Top and bottom panels show the galaxies from the S5 and V5 volumes, respectively. The {\apo} and AUGIRA results are plotted using blue and red colours, respectively. The solid (dashed) vertical lines mark the median $j$ at $z = z_{\rm acc}$ ($z = z_{\rm sf}$). At $z = z_{\rm acc}$, {\apo} and {\auriga} simulations exhibit relatively similar distributions as a result of the similar tidal torquing. In contrast, at $z = z_{\rm sf}$, the {\apo} runs tend to have lower $j$ compared to the {\auriga} runs, suggesting that their stellar feedback subgrid models affect the gas sAM evolution differently.
        % \zeng{\textit{[For the convenience of readers, I suggest adding a legend about solid and dashed lines to this figure.]}}
        }
	%\caption{The specific angular momentum magnitude $j$ distribution of each tracer we selected at $z_{\rm{acc}}$ and $z=z_{\rm{sf}}$ in {\auriga} (red) and {\apo} (blue). The vertical dashed (solid) line to represent the median $j$ at $z=z_{\rm{sf}}$ (halo accretion $z_{\rm{acc}}$).}
    \label{fig:Total}
\end{figure*}

We track the in-situ star particles within $0.1 R_{200}$ of the $z=0$ galaxies back to $z = z_{\rm acc}$, the time when the gas tracers first enter the main halo progenitor, and plot the probability distribution functions (PDFs) of the magnitude of their sAM, $\log_{10} j$, as filled histograms in Figure~\ref{fig:Total}. The median sAM of each distribution is marked using a vertical solid line. Remarkably, the distributions from the the two simulations are relatively similar at $z = z_{\rm{acc}}$. This outcome arises from the similar large-scale tidal torquing, as these simulations start from the same initial conditions, which determines the evolution of angular momentum in the early stage \citep[e.g.][]{1969ApJ...155..393P, 1970Afz.....6..581D,1984ApJ...286...38W, 2015MNRAS.449.2087D}.

% \zeng{\textit{[Just curious: what does the distribution of circularities, $\epsilon$, look like at $z=z_{\mathrm{acc}}$? Will the distributions from the two simulations also be similar?]}}

As a comparison, we plot the $j$ distributions at the endpoint of the gas evolution, $z = z_{\rm sf}$ (i.e. the moment when the gas tracers are converted into star particles) in Figure~\ref{fig:Total} (unfilled histograms). Once again, the median sAM of each distribution is denoted by a vertical dashed line. In contrast to the distributions at $z = z_{\rm acc}$, notable differences emerge between the {\apo} and {\auriga} simulations, with {\auriga} tracers exhibiting higher $j$ values. From $z_{\rm acc}$ to $z_{\rm sf}$, the average sAM loss amounts to approximately 0.70 (0.97) dex in the {\auriga} ({\apo}) simulations.

This comparison hints that the difference in the final stellar sAM between the two simulations mainly originates from the processes within haloes, rather than from initial accretion. After being accreted into the main halo progenitor and before being converted into stars, gas can experience substantial changes in angular momentum due to the energy and momentum received from feedback processes. The observed disparity in sAM distributions at $z = z_{\rm sf}$ suggests that different subgrid models for these physical processes may lead to distinct sAM evolution in these simulations.

\subsection{Distributions of characteristic time}\label{subsec:cdf_time}

\begin{figure*}
	\includegraphics[width=\textwidth]{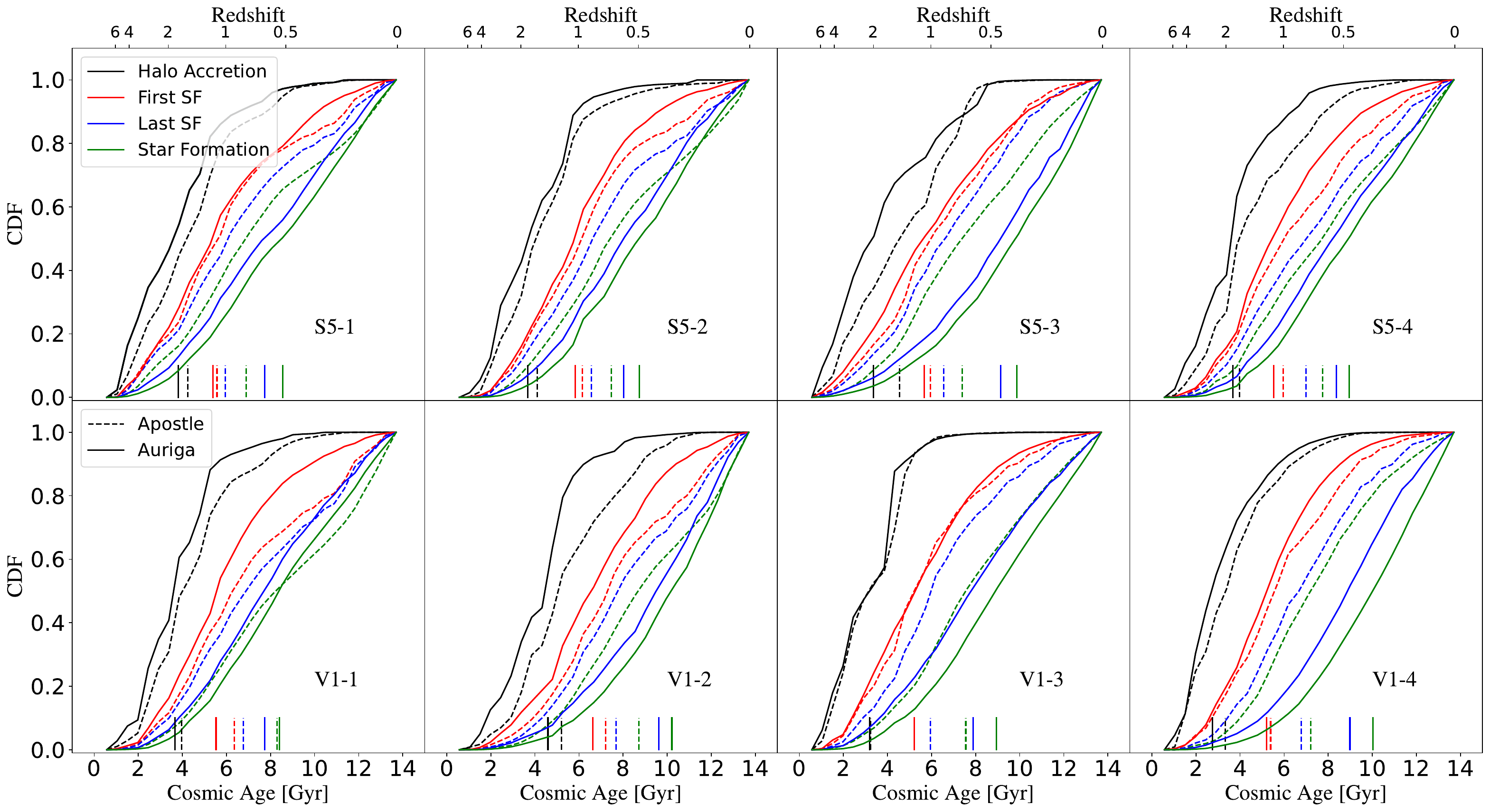}
        \caption{Cumulative distribution functions of different characteristic event time. The upper (lower) panels display the galaxy pair in the S5 (V5) volume. The halo accretion time, first star-forming time, last star-forming time, and star formation time are plotted using black, red, blue, and green lines, respectively. Solid and dashed lines distinguish the {\auriga} and {\apo} simulations. The vertical line segments denote the median time of different events. The top $x$-axis shows redshift, while the bottom $x$-axis displays the cosmic age.}
	
    \label{fig:Time}
\end{figure*}

We employ some characteristic events which defined in Section~\ref{subsec:chara_events} to delineate the entire evolution history. To quantify the typical values for the characteristic time, we plot their cumulative distribution functions (CDFs) in Figure~\ref{fig:Time}. The {\auriga} and {\apo} runs are distinguished by solid and dashed lines, respectively. The median time for each CDF is marked by a vertical line segment.

For example, in the S5-1 galaxy pair, the distributions of halo accretion time (black) are relatively similar in the two simulations, with the median halo accretion time being $t_{\rm{acc}}^{\rm Au} = 3.82$ Gyr and $t_{\rm{acc}}^{\rm Ap} = 4.25$ Gyr (where $t$ denotes the cosmic age) for {\auriga} and {\apo}. For the first star-forming time (red), the median values for {\auriga} and {\apo} being $t_{\rm fsf}^{\rm Au} = 5.38$ Gyr and $t_{\rm fsf}^{\rm Ap} = 5.57$ Gyr, respectively. However, the distributions for the last star-forming time (blue) and the star formation time (green) show significant differences between the two simulations. In {\auriga}, the median last star-forming time is $t_{\rm lsf}^{\rm Au} = 7.74$ Gyr and the median star formation time is $t_{\rm sf}^{\rm Au} = 8.56$ Gyr. In contrast, in {\apo}, these median values are $t_{\rm lsf}^{\rm Ap} = 5.95$ Gyr and $t_{\rm sf}^{\rm Ap} =  6.89$ Gyr. 

Similar trends are observed for other galaxy pairs. The substantial differences in the CDFs of $z_{\rm lsf}$ and $z_{\rm sf}$ between the two simulations indicate that the evolution of gas tracers after $z = z_{\rm fsf}$ differs markedly in {\auriga} and {\apo}, with gas taking more time to form stars after crossing the star-forming threshold in {\auriga}. As we will detail in Section~\ref{subsec:recycle}, this difference relates to the distinct behaviours of recycled fountain flows in these two simulations.

\subsection{Evolution of angular momentum from $z=z_{\rm acc}$ to $z=0$}

\begin{figure}
	\includegraphics[width=\linewidth]{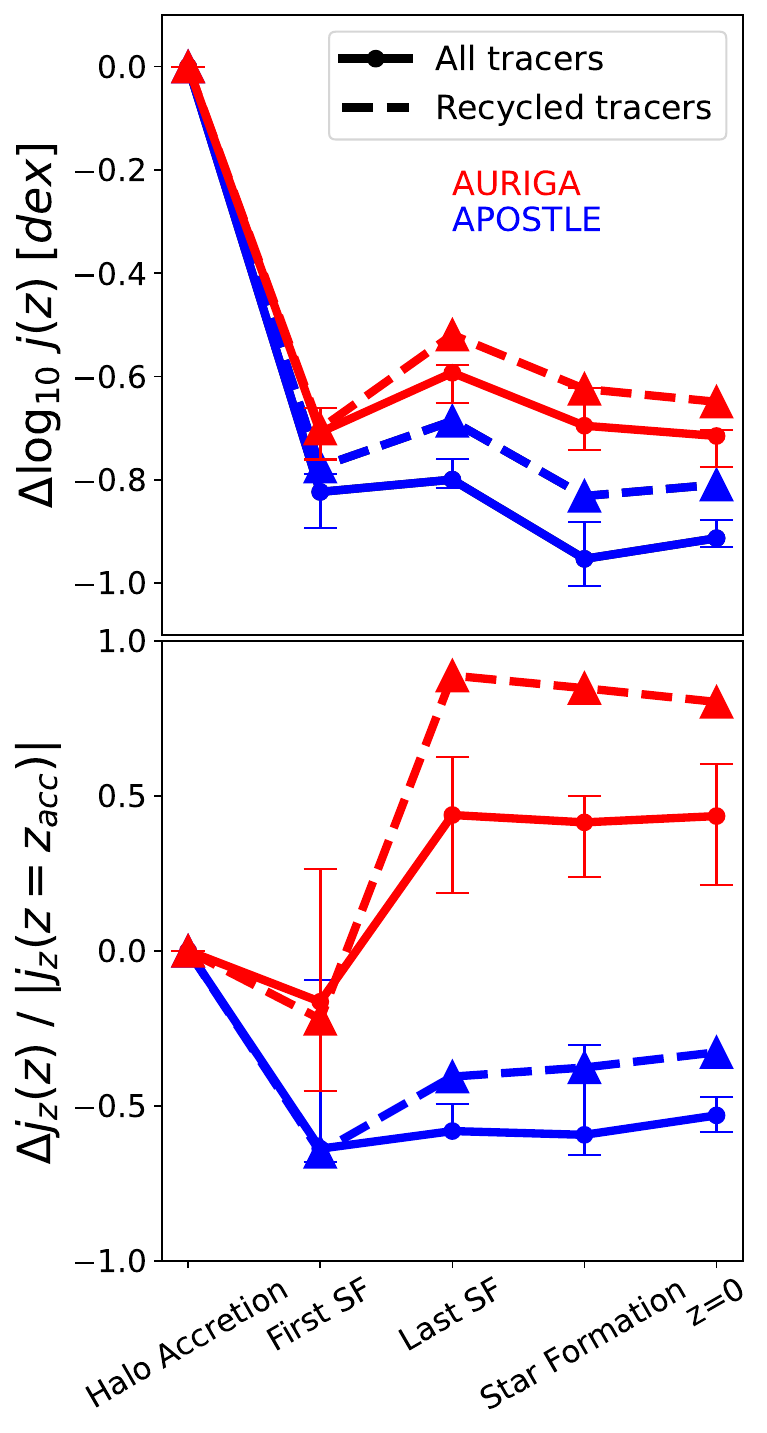}
        \caption{Evolution of angular momentum. {\it Top:} The loss of sAM relative to the angular momentum at halo accretion, $\Delta \log_{10} j(z)$. Markers represent the median sAM loss of all eight galaxies, with error bars indicating the 16th to 84th percentiles. Results from the {\auriga} and {\apo} simulations are plotted in red and blue, respectively. Solid and dashed lines show the results computed from all tracers and only tracers which have experienced the baryon cycling process (i.e. recycled tracers), respectively. For clarity, only the scatters for the `all tracers' case are shown, as the `recycled tracers' case exhibits similar scatters. {\it Bottom:} Similar to the top panel, but for the loss of the $z$-component of sAM, $\Delta j_{z}(z) / |j_{z}(z_{\rm acc})|$.}
    \label{fig:AM_SF_ind_med}
\end{figure}

To study the entire evolution history of angular momentum, we compute the sAM of each tracer at different characteristic events defined in Section~\ref{subsec:chara_events}. As we have shown in Section~\ref{subsec:overall_evo_am}, the distributions of sAM at $z = z_{\rm acc}$ from the two simulations are relatively similar. Therefore, in this subsection, we focus on the sAM change relative to the sAM at $z = z_{\rm acc}$. 

Specifically, for each tracer in a galaxy, we compute the sAM loss with respect to its $j_i$ at $z = z_{\rm acc}$, i.e. $\Delta \log_{10} j_i (z) = \log_{10} j_i (z) - \log_{10} j_i (z_{\rm acc})$. The median sAM loss of this galaxy is then defined as the median $\Delta \log_{10} j (z)$ of all tracers. In Figure~\ref{fig:AM_SF_ind_med}, the median sAM loss computed over eight galaxies is plotted using solid connected lines in the top panel, with the error bar representing the 16th to 84th percentiles. In addition, to examine the evolution of the $z$-axis component of sAM (i.e. the component along the direction of the stellar sAM at the present day $z = 0$), we plot the median relative loss of $j_z$ normalized to $|j_z|$ at the halo accretion event (solid connected lines) in the bottom panel.

We can observe the following evolution process of angular momentum from Figure~\ref{fig:AM_SF_ind_med}:

\begin{description}

\item {\it From $z=z_{\rm acc}$ to $z=z_{\rm fsf}$}: After being accreted into dark matter haloes, each gas tracers in {\auriga} ({\apo}) on average lose approximately 0.71 (0.82) dex of their sAM magnitudes when entering the central galactic disc and crossing the star-forming density threshold for the first time (i.e. at $z = z_{\rm fsf}$). During this evolution, tracers lose ${\sim}16 ({\sim}64)$ per cent of $j_z$ in {\auriga} ({\apo}). As pointed out by \citet{2017ApJ...841...16D}, the decrease in these tracers' sAM results from the transport of angular momentum to other components (e.g. dark matter and baryonic) and the cancellation with other baryons which have been accreted with different sAM directions. The larger sAM loss in the {\apo} simulations implies that the angular momentum transfer and cancellation are more effective in the {\textsc{eagle}} model.

\item {\it From $z = z_{\rm fsf}$ to $z = z_{\rm lsf}$}: The average sAM magnitude of gas tracers in {\auriga} increases by ${\sim} 0.12$ dex, and the average increase in $j_z$ is ${\sim} 60$ per cent. In contrast, the average sAM magnitude in {\apo} remains almost constant. As we will detail in Section~\ref{subsec:recycle}, this process is related to the so-called galactic fountain, where gas flows are ejected by stellar feedback and recycled back, acquiring angular momentum through mixing with circumgalactic medium (CGM) and newly accreted gas \citep[e.g.][]{2012MNRAS.419..771B,2019MNRAS.490.4786G}. The almost constant sAM in this phase in {\apo} reflects that its recycled fountain is significantly weaker compared to {\auriga}.

\item {\it From $z = z_{\rm lsf}$ to $z = z_{\rm sf}$}: The average sAM magnitudes of gas tracers in both simulations decrease by ${\sim} 0.1$ dex, while the average $j_z$ is almost unchanged. This indicates that during this stage, as the gas tracers continue to cool, condense into the galactic disc, and finally turn into stars, the sAM components perpendicular to the stellar disc direction are gradually dissipated into the interstellar medium (ISM), leaving only the sAM component along the disc direction.

\item {\it From $z = z_{\rm sf}$ to $z = 0$}: The stellar sAM are almost conserved in both simulations.

\end{description}

To summarize, by tracing the evolution history of the star particles in a $z = 0$ disc galaxy, we find that the sAM difference at $z = 0$ between the {\apo} and {\auriga} simulations mainly develops between the halo accretion and last star-forming phases, which is closely related to the galactic fountain. In the following subsection, we delve into more details of the effects of fountain flows.

\bigskip\bigskip

\subsection{Effect of recycled fountain flows}\label{subsec:recycle}

\begin{figure}
	\includegraphics[width=\linewidth]{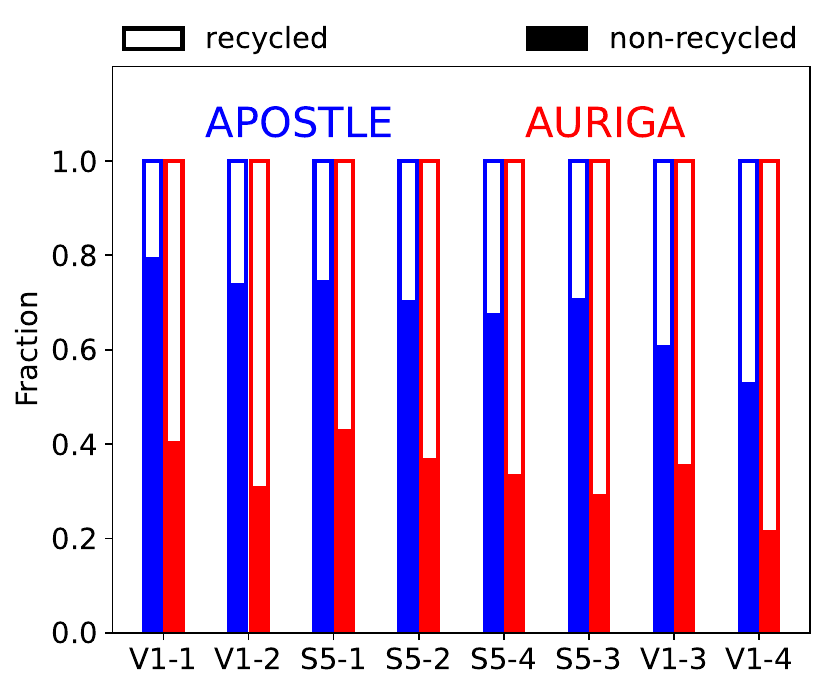}
        \caption{Number fractions of recycled (unfilled) and non-recycled (filled) tracers in different {\auriga} (red) and {\apo} (blue) galaxies. From left to right, the galaxy pairs are sorted in descending order by virial mass. {\apo} galaxies exhibit higher fractions of non-recycled gas tracers, while {\auriga} galaxies are dominated by recycled gas tracers.}
    \label{fig:fraction}
\end{figure}

\begin{figure*}
	\includegraphics[width=0.8\linewidth]{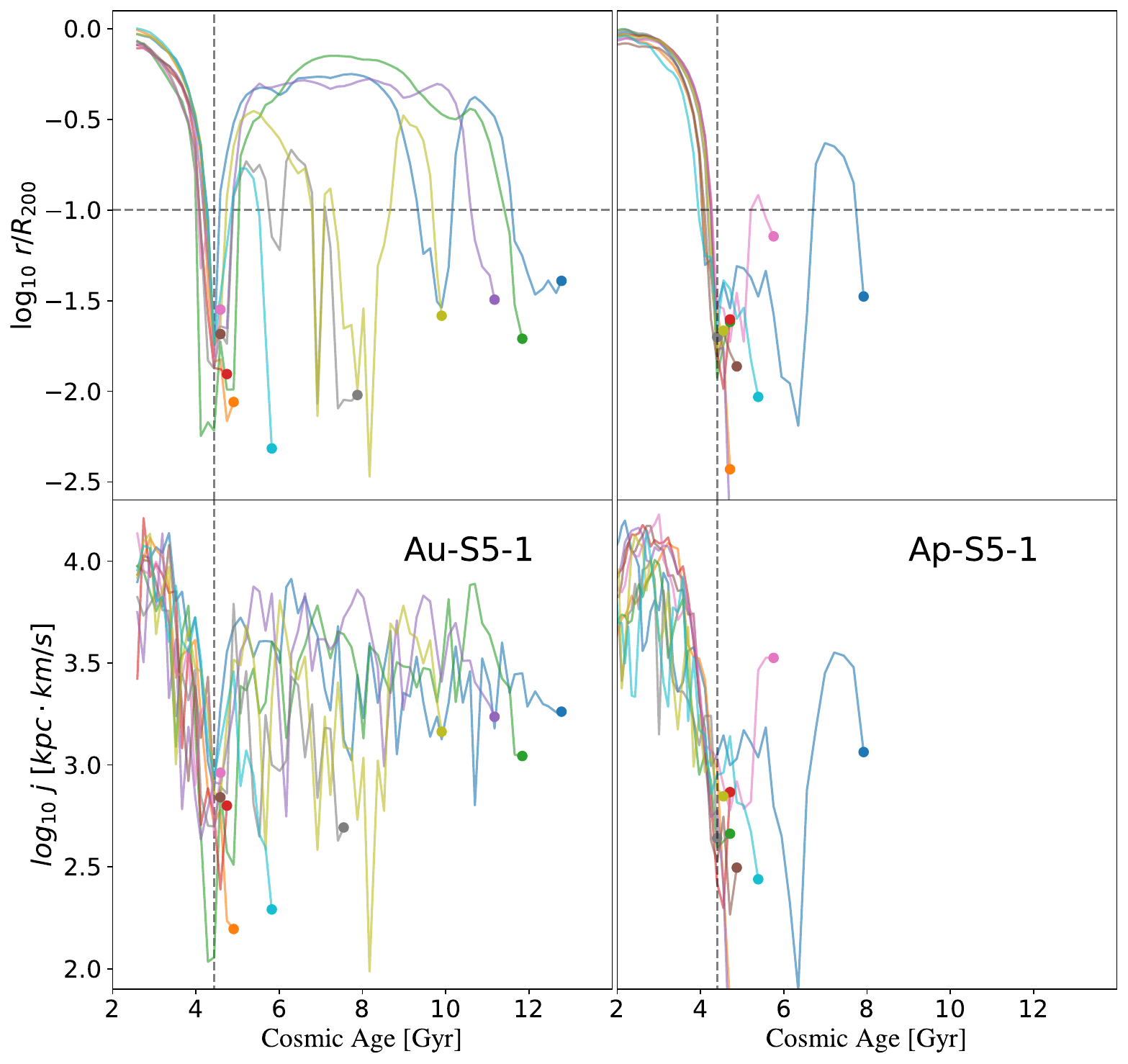}
        \caption{Evolution of ten randomly selected gas tracers with the same first star-forming time (marked by the vertical dashed lines) in the Au-S5-1 (left) and Ap-S5-1 (right) galaxies. Top panels show the rescaled distance from the gas tracer to the galaxy centre, and bottom panels plot the sAM of gas tracers. The horizontal dashed line marks the radius of $0.1R_{200}$. Different gas tracers in a simulation are distinguished by different colours. The evolution trajectories are plotted until the gas tracers are converted into stars (marked by the filled circles). Compared to {\apo}, more gas tracers (which eventually form stars) in {\auriga} experience the recycling process. When the outflow-recycled gas is ejected by stellar feedback, it acquires angular momentum through mixing with the high angular momentum CGM and settles in the relatively outer disc to form stars.}
    \label{fig:traject}
\end{figure*}

Previous studies have shown that after being ejected from the ISM, the fountain flows mix with the low-density, high-temperature CGM gas, which usually has high angular momentum. As a result, galactic fountains effectively redistribute angular momentum, with the recycling gas acquiring angular momentum from the CGM \citep[e.g.][]{2012MNRAS.419..771B, 2017MNRAS.470.3167V, 2019MNRAS.490.4786G}. This mechanism can explain the increase of sAM for gas tracers between $z = z_{\rm fsf}$ and $z = z_{\rm lsf}$ in simulations.

To examine the above scenario, we quantify the fraction of recycled flows in the gas tracers. Specifically, we record the number of times each gas tracer leaves star-forming state after $z = z_{\rm fsf}$. Based on whether they leave the star-forming state at least once, gas tracers are divided into non-recycled and recycled categories.\footnote{Note that here, for simplicity, we do not apply a radial velocity cut when defining recycled fountain flows. As shown in figure B2 of \citet{2020MNRAS.497.4495M}, the radial velocity cut has only a minor effect on recycled gas accretion rates.} The number fractions of recycled and non-recycled tracers for each galaxy pair are summarized in Figure \ref{fig:fraction}. Note that in this figure, the galaxy pairs are plotted from left to right in descending order by virial mass.

In {\apo}, the majority of tracers are categorized as non-recycled gas, with the fractions of non-recycled gas being higher in more massive galaxies. Specifically, the fraction of non-recycled gas is ${\sim} 80$ per cent for Milky Way-sized galaxies while it is ${\sim} 50$ per cent for dwarf galaxies. In contrast, the majority of tracers in {\auriga} are recycled gas, with the fraction of non-recycled (recycled) gas being ${\sim} 35$ $({\sim} 65)$ per cent for all galaxies. The fractions of non-recycled or recycled gas do not show a clear mass dependence in {\auriga}. 

This comparison highlights the differences in galactic fountains between the two simulations, which are a direct outcome of their different stellar feedback subgrid models. Previous studies \citep[e.g.][]{2022MNRAS.514.3113K, Wright2024} have shown that in galaxies below ${\sim} 10^{12}~{\rm M}_{\sun}$, the {\textsc{eagle}} stellar feedback model tends to eject gas particles from the ISM to several virial radii, reducing recycling flows within the virial radius and impeding cosmic gas accretion. In contrast, in the {\auriga} simulations, the ejected gas flows driven by stellar feedback typically stall before reaching the galaxy's virial radius, resulting in a higher degree of recycled fountain flows. As we focus on gas that eventually forms stars, in {\apo}, this gas usually does not experience strong ejections from stellar feedback. In {\auriga}, however, even gas ejected by stellar feedback can recycle back to the galactic disc and form stars.

To illustrate the recycled flow, we randomly select ten gas tracers with the same first star-forming time, $z_{\rm fsf}$, from Galaxy S5-1 in both simulations. The time evolution of their distances to the galaxy centre (rescaled by the virial radius, i.e. $r/R_{200}$) and sAM before being converted into stars are plotted in Figure~\ref{fig:traject}. In Au-S5-1, the majority of the selected gas tracers (6 over 10) are ejected to the CGM beyond $0.1 R_{200}$ and then recycle back to form stars. These outflow-recycled gas tracers, which have low angular momenta at $z = z_{\rm fsf}$ \citep[agreeing with][]{2011MNRAS.415.1051B}, acquire angular momenta through mixing with the high angular momentum CGM gas, eventually settling in the relatively outer disc. This agrees with the findings of previous studies \citep[e.g.][]{2012MNRAS.419..771B,2019MNRAS.490.4786G}. In Ap-S5-1, the majority of the selected gas tracers form stars quickly after $z = z_{\rm fsf}$, with only 2 over 10 being weakly ejected outside the galactic disc and forming stars shortly after recycling back. In addition, we notice that after $z_{\rm fsf}$, the recycled gas tracers in {\auriga} take more time to evolve into stars compared to those in {\apo}, providing an explanation for the observed differences in the CDFs of characteristic time in Section~\ref{subsec:cdf_time}.

Finally, to investigate directly the effect of recycled fountain flows on angular momentum evolution, we consider only the recycled gas tracers and plot their angular momentum evolution in Figure~\ref{fig:AM_SF_ind_med} using dashed lines. We can clearly see that between $z = z_{\rm fsf}$ and $z = z_{\rm lsf}$, the increase in sAM in {\auriga} is larger compared to the `all tracers' case. Intriguingly, we also observe an increase in sAM in {\apo}, reflecting that the recycled flows can indeed boost the angular momentum regardless of simulations. The almost constant sAM in the `all tracers' case of {\apo} is a result of the dominant fraction of non-recycled tracers.

\section{Summary}\label{sec:summary}

In this paper, we have studied the impact of stellar feedback subgrid models on the evolution of disc galaxies' angular momenta by using two suits of zoom-in simulations, {\apo} and {\auriga}. By tracing the evolution history of the $z = 0$ star particles from a Lagrangian perspective, we characterize the evolution of the particle tracers' angular momenta at five different events: halo accretion, first star-forming, last star-forming, star formation, and the present state. We summarize the following:

\begin{enumerate}
    \item At $z = 0$, both the {\apo} and {\auriga} disc galaxies closely follow the observed stellar angular momentum--mass relation. However, the {\auriga} galaxies exhibit slightly higher sAM and higher stellar masses compared to their {\apo} counterparts (Figure~\ref{fig:jM}).
    \item By tracing the Lagrangian tracers of the in-situ star particles in $z = 0$ galaxies back to the halo accretion time, we find that the sAM distributions of the two simulations are relatively similar. This suggests that the main driver of the present-day difference in sAM is not initial accretion but rather the later processes occurring inside haloes (Figure~\ref{fig:Total}).
    \item After being accreted into dark matter haloes ($z = z_{\rm acc}$),  gas tracers in {\auriga} ({\apo}) on average lose ${\sim} 0.71({\sim} 0.82)$ dex of their sAM when entering the central galactic disc and reaching the star-forming condition for the first time ($z = z_{\rm fsf}$). From $z = z_{\rm fsf}$ to $z = z_{\rm lsf}$, the average sAM of gas tracers in {\auriga} increases by ${\sim}0.12$ dex, while it remains almost constant in {\apo}. In the remaining evolution stages (i.e. last star-forming--star formation--present-day), the average angular momentum evolution of the Lagrangian tracers is similar in the two simulations (Figure~\ref{fig:AM_SF_ind_med}).
    \item Different implementations of stellar feedback in the two simulations lead to distinct recycled fountain flows, which further affect the angular momentum evolution history. {\auriga} galaxies contain a higher fraction (on average ${\sim} 65$ per cent) of recycled gas tracers, and this gas acquires angular momentum through mixing with the high angular momentum CGM. In contrast, the Lagrangian tracers of the $z=0$ {\apo} stellar discs are dominated by non-recycled gas (i.e. up to ${\sim} 80$ per cent for Milky Way-sized galaxies) and the acquisition of angular momentum from the CGM is negligible (Figures~\ref{fig:fraction} and \ref{fig:traject}). As a result, {\auriga} disc galaxies exhibit higher angular momentum at $z = 0$ compared to {\apo} galaxies.
\end{enumerate}

Our results reveal the different angular momentum evolution histories of disc galaxies in the {\apo} and {\auriga} galaxy formation subgrid models. Future investigations into the impact of baryonic physics subgrid models (such as stellar feedback) on the masses and sizes of stellar discs will be valuable, as they will further help to understand the origin of differences in the disc size--halo spin relation observed in various hydrodynamical simulations \citep{2023MNRAS.518.5253Y}.

%%%%%\bigskip\bigskip

\section*{Acknowledgements}

We would like to thank Simon D. M. White for useful discussions and advice. We also thank the anonymous referee for their helpful comments. We acknowledge the supports from the National Natural Science Foundation of China (Grant No. 11988101) and the K. C. Wong Education Foundation. SS acknowledges support from NSFC grant (No. 12273053), CAS Project for Young Scientists in Basic Research Grant (No. YSBR-062), and the science research grants from the China Manned Space Project with No. CMS-CSST-2021-B03. AF is supported by a UKRI Future Leaders Fellowship (grant no MR/T042362/1). RJW acknowledges the supports by the Forrest Research Foundation and the European Research Council via ERC Consolidator Grant KETJU (no. 818930). 

This work used the DiRAC@Durham facility managed by the Institute for Computational Cosmology on behalf of the STFC DiRAC HPC Facility (www.dirac.ac.uk). The equipment was funded by BEIS capital funding via STFC capital grants ST/K00042X/1, ST/P002293/1, ST/R002371/1, and ST/S002502/1, Durham University and STFC operations grant ST/R000832/1. DiRAC is part of the National e-Infrastructure.

We thank the developers of the open-source {\sc python} packages that were used in the data analysis of this work, including {\sc matplotlib} \citep{Hunter2007}, {\sc numpy} \citep{Harris2020}, {\sc scipy} \citep{virtanen2020scipy} and {\sc h5py} \citep{2020zndo...4250762C}.

%%%%%%%%%%%%%%%%%%%%%%%%%%%%%%%%%%%%%%%%%%%%%%%%%%
\section*{Data Availability}

The data underlying this article will be shared on reasonable request to the corresponding author.
%The other data directly related to this publication and its figures is available on request from the corresponding author.



\bibliographystyle{mnras}
\bibliography{example} % if your bibtex file is called example.bib

\bsp	% typesetting comment
\label{lastpage}
\end{document}